\documentclass[prd,superscriptaddress,a4paper,showpacs,showkeys,nofootinbib]{revtex4-1}
\usepackage{graphicx}
\usepackage{xcolor}
\usepackage{amsmath}
\usepackage{subfigure}
\usepackage{multirow}
\usepackage[hidelinks]{hyperref}
\def\equationautorefname#1#2\null{%
  Eq.\;(#2\null)%
}
\def\figureautorefname#1\null{%
  Fig.#1\null
}

\begin{document}

\begin{flushright}
MS-TP-22-52
\end{flushright}

\title{Fundamental-mode eigenfrequencies of neutral and charged twin neutron stars}
\author{Victor P. {\sc Gon\c{c}alves}}
\email{barros@ufpel.edu.br}
\affiliation{Institute of Physics and Mathematics, Federal University of Pelotas \\
  Postal Code 354,  96010-900, Pelotas, RS, Brazil}
\affiliation{Institute of Modern Physics, Chinese Academy of Sciences,
  Lanzhou 730000, China}
\affiliation{Institut f\"ur Theoretische Physik, Westf\"alische Wilhelms-Universit\"at M\"unster,
Wilhelm-Klemm-Straße 9, D-48149 M\"unster, Germany
}

\author{Jos\'e C. {\sc Jim{\'e}nez}}
\email{josec.jimenez91@gmail.com}
\affiliation{Instituto de F\'{\i}sica, Universidade de S\~ao Paulo,\\
Rua do Mat\~ao 1371, 05508-090 S\~ao Paulo-SP, Brazil}

\author{Lucas {\sc Lazzari}}
\email{lucas.lazzari@ufpel.edu.br}
\affiliation{Institute of Physics and Mathematics, Federal University of Pelotas \\
  Postal Code 354,  96010-900, Pelotas, RS, Brazil}

\begin{abstract}
  We investigate the effects of rapid and slow conversions on the fundamental-mode eigenfrequencies of hybrid neutron stars having highly discontinuous transitions between hadronic and quark matter, the so-called twin stars. We analyze some characteristic cases in the available parameter space of the equations of state for the hadronic and quark phases. Furthermore, we also consider the possibility that these stellar configurations are electrically charged. Our results indicate that for neutral configurations under rapid conversions the stability window coincides with the usual stability criterion, i.e. $\partial M/\partial p_c > 0$ and that the two branches are disconnected. This discontinuity remains when electric charge is considered, but the usual criterion is not sufficient to determine the star's stability. On the other hand, slow conversions connect initially disconnected branches and increase the stability window of the hybrid configurations. For both conversion speeds, the presence of electric charge diminishes the magnitude of the eigenfrequencies and its stability window. 
\end{abstract}

\maketitle

\section{Introduction}
\label{sec:intro}

The advent of multimessenger astronomy in recent years has opened a new window to look more carefully into the underlying laws governing astrophysical objects. For instance, black-hole (BH) mergers were directly observed by the LIGO/Virgo collaborations through their gravitational-wave emissions~\cite{LIGOScientific:2016aoc}. Besides, NICER gives us the opportunity to measure X-ray emissions from neutron stars (NS) in order to obtain their masses and radii with unprecedented precision~\cite{Miller:2019cac,Riley:2019yda}. Remarkably, NS mergers sent us both kinds of emissions plus the potential opportunity to obtain neutrino data~\cite{Meszaros:2019xej}. These measurements (except pure BH situations) have been used to constrain the equation of state (EoS) for cold and dense strongly interacting matter present in NS interiors~\cite{Annala:2017llu,TheLIGOScientific:2017qsa}. This kind of procedure becomes specially important at intermediate densities since first-principle calculations through in-medium quantum chromodynamics (QCD) are still unable to yield reliable results even using lattice QCD calculations due to the sign problem~\cite{deForcrand:2010ys}.

At high densities, one expects a confinement/deconfinement transition among hadronic and quark matter (QM)~\cite{Annala:2019puf}. This QCD transition can be manifested in the static structure and stability of the so-called hybrid NS which contains a QM core surrounded by a hadronic mantle~\cite{Glendenning:2000}. Of course, this kind of transition could be especially realized in heavy NSs {since several recent and robust studies \cite{Annala:2019puf, Kojo:2020krb, Komoltsev:2021jzg,Annala:2021gom,Pang:2022rzc} suggest that their stellar cores support densities around several times the nuclear saturation density, $n_{0}$, i.e. usually in the range $\sim 5-10~n_{0}$.} 

Thermodynamically, the transition between phases might be soft (crossover) or discontinuous (first-order)\footnote{Certainly, the same reasoning applies in NS mergers although thermal effects should be added into the EoS in order to account for the heating when the stars collide~\cite{Kiuchi:2009jt}.}. For the former case, many studies~\cite{Sahu:2001iv,Gupta:2002fk,Annala:2019puf,Jimenez:2021wil} suggested small variations in the mass-radius diagram which even in some cases become negligible  due to the high degree of softness of the EoS. {Of course, depending on the hadron/quark model considered, one has that with reasonable values of their free parameters, one can still fulfill the two-solar mass limit by means of a crossover transition \cite{Kapusta:2021ney,Kojo:2021wax}. We do not consider this interesting possibility here.}

{Moreover, several} calculations indicate that when the transition is strongly discontinuous, an ultra-dense disconnected new branch of compact objects arise, the so-called twin (hybrid) NSs, that are considerably composed by QM in their cores~\cite{Dexheimer:2014pea,Alvarez-Castillo:2017qki,Schertler:2000xq,Christian:2020xwz}. In this sense, this kind of exotic objects serve to probe the effects of a first-order transition {on the radial oscillation frequencies, which are tightly related to the dynamical stability of the stellar configurations. From the observational point of view, radial and non-radial modes are activated in static and rotating compact stars through the accretion of matter \cite{Passamonti:2005ac} producing seismic motions \cite{Chugunov:2006kk} from asteroids and comets \cite{Geng:2015vza}, planet/compact-star binaries \cite{Kutschera:2020gxx}, and in protoneutron stars \cite{Passamonti:2007tm}. It should be noted that from these astrophysics scenarios we are immediately led to consider the possibility of charged compact stars since external agents, like comets or companion planets, increase electric charge in the assumed neutral stars along its lifetime. In other words, very old NS might be neutral but young/adult ones could still carry some large amount of electric charge when observed on Earth.} 
{Unfortunately, until now it is a challenge to have reliable and direct measurements of the radial modes due to large uncertainty, for instance, associated to atmospheric models of the NS surface. However, it was proven in Refs.~\cite{Passamonti:2005cz,Passamonti:2007tm}
that these radial modes can be inferred indirectly through the direct measurement of non-radial modes, e.g. axial and polar modes, producing gravitational waves that could be potentially measurable in next generation detectors. Besides, knowledge of the the dynamical stability of the stellar configurations serve to constrain the degree of discontinuity of the EoS allowed to obtain stars which can be realized in nature~\cite{Jimenez:2021wil} and, more importantly for this work, twin stars}\footnote{{We note that a recent study performed in Ref.~\cite{Espino:2021adh} pointed out that} {it may be difficult to directly form stable twin stars if the phase transition is sustained over a large jump in energy density and that stable twin stars can be formed from a massive hybrid star that loses mass and enters the stable twin star regime. Such results have been derived considering a particular set of EOSs and assuming just a rapid conversion process from hadronic to QM. Surely, the extension of the results obtained in Ref.~\cite{Espino:2021adh} for the EoSs, density jumps and slow conversions considered in this paper is a theme that deserves a detailed study in the future.}}, {which have strongly discontinuous transitions between confined and deconfined matter.}

{Although it is simple to model the aformentioned transition from a thermodynamic viewpoint}, the microphysics of the phase conversion dynamics is poorly understood since ab initio attempts fail due to the nonperturbative nature of QCD near the transition point. Thus, many studies concerning this issue are fully based upon phenomenological results with model-dependent results~\cite{Bombaci:2016xuj,Voskresensky:2002hu,Endo:2011em,Wu:2018zoe}. However, one can under- and overestimate the timescale of the conversion process from hadronic to QM and deduce the stellar modifications which in principle could be measurable~\cite{Pereira:2017rmp,Tonetto:2020bie,Pereira:2020cmv,Lugones:2021bkm,Lugones:2021zsg}.  In particular, the results obtained in Ref. \cite{Pereira:2017rmp}  indicated that the dynamical  stability  of  hybrid  stars  is  strongly  influenced  by  the speed  of  the  phase  conversion.  For rapid conversions, where the conversion time between both phases is much smaller than the characteristic time of oscillations, one has that the usual stability criterion, i.e. $\partial M/ \partial p_{c} \geq 0$, is enough to determine the stability of hybrid stars. On the other hand, when the conversion time is much larger than that of the oscillation (slow conversions), such criterion has been shown to be insufficient to determine the stability of hybrid stars with one or two sharp phase transitions~\cite{Pereira:2017rmp,Lugones:2021bkm,Lugones:2021zsg,Goncalves:2022ymr}.

In addition, {and as already briefly discussed}, in recent years several studies have analyzed the impact of electric charge distribution on the structure properties of compact stars~\cite{Brillante:2014lwa,arbanil2015,Panotopoulos:2019wsy,Goncalves:2020joq,Panotopoulos:2020hkb,Jasim:2021cft,Goncalves:2021pmr}. {In principle, {apart from mass accretion}, a compact star could {also} be electrically charged after the merger of two NSs if the remnant is stable. Although the existence of charged compact stars is still a theme of debate, the results derived in the quoted references indicate that these objects can be stable and have properties that differ {sizeably} from its neutral counterpart}. In particular, in Ref.~\cite{Goncalves:2021pmr}, we { have shown  that charged twin hybrid stars are more massive than their neutral counterparts, with the presence of electric charge diminishing the stability region but generating stable configurations with large masses ($ \gtrsim {2.2}{M_{\odot}}$), in agreement with recent measurements of heavy NS masses~\cite{Fonseca:2016tux,Linares:2018ppq,Cromartie:2019kug}. More interestingly, for some amounts of electric charge the stellar masses even surpass the $2.5 M_{\odot}$ limit put by the unknown object detected by the GW190814 event~\cite{LIGOScientific:2020zkf}. {However, the stability analysis in Ref.~\cite{Goncalves:2021pmr} was performed assuming the usual stability criterion, which is only true for rapid conversions in the neutral scenario, as will be shown later on.}

Our goal in this paper is to perform a systematic study of neutral and charged twin stars considering rapid and slow conversions and estimate the impacts of electric charge and conversion speeds on the eigenfrequency of the fundamental mode. As a consequence, we will be able to analyze the stability of these stars for the cases where  $\partial M/ \partial p_{c} \geq 0$ is insufficient to determine the stability of hybrid stars. In this sense, the present study complements the analysis performed in Ref.~\cite{Goncalves:2021pmr}. As in Ref.~\cite{Goncalves:2021pmr}, the QM phase of the neutral and charged twin stars will be modelled assuming two distinct models for the EoSs (the constant speed of sound (CSS) parametrization~\cite{Alford:2004pf,Alford:2013aca} and the multipolytropes of Ref.~\cite{Alvarez-Castillo:2017qki}) and the eigenfrequencies will be estimated assuming that the phase conversion is slow or rapid. As we will demonstrate below, the magnitude of the eigenfrequencies and the stability window are sensitive to the treatment of the phase conversion and to the magnitude of electric charge in the star.

This paper is organized as follows. In the next section, we will present a brief review of the formalism needed to describe radial oscillations as well as rapid and slow conversions in hybrid stars. Moreover, we describe the distinct models of EoSs used for the hadronic and quark phases and present the model assumed for the electric charge distribution. In Sec.~\ref{sec:results}, the predictions for the fundamental mode eigenfrequencies, derived assuming distinct EoSs and different treatments for the phase conversion, are presented. Finally, in Sec.~\ref{sec:conclusion} we summarize our main results and conclusions.

\section{Formalism}
\label{sec:setup}

Although the effects of rapid and slow conversions in first-order phase transitions are not yet fully understood microscopically~\cite{Bombaci:2016xuj,Voskresensky:2002hu,Endo:2011em,Wu:2018zoe}, one can still study somewhat quantitatively their influence on the dynamical stability of hybrid NSs in hydrostatic equilibrium. See Ref.~\cite{Haensel:1989wax} for related studies in the Newtonian approximation and Ref.~\cite{Pereira:2017rmp} within general relativity. In {the latter}, it was proven that the radial oscillation equations of Gondek-Rosińska {\it et al.}~\cite{Gondek:1997fd} are not directly affected by the discontinuity at transition between phases but they influence only the usual boundary conditions\footnote{The same reasoning applies for the Sturm-Liouville problem derived by S. Chandrasekhar although obviously somewhat more involved theoretically but far more complicated numerically.} on the Lagrangian displacements of the relative radial coordinate, $\xi=\Delta{r}/r$, and pressure, $\Delta{p}$. In what follows, we summarize the main aspects behind this formalism.

Historically, after S. Chandrasekhar~\cite{Chandrasekhar:1964zza} derived the equations that determine the stability of one-phase compact stars, one was able to discriminate realistic from unstable solutions of his equations through analyzing the behavior of the squared eigenfrequency of the fundamental mode, i.e. $\omega^{2}_{n=0}$. When $\omega^{2}_{n=0} \ge 0$ one has harmonically oscillating configurations, while for $\omega^{2}_{n=0} < 0$, the solution is unstable, with the star collapsing to become a black hole. In recent years, several authors have demonstrated that it is numerically advantageous to transform the  original second-order differential equation in two first-order  equations for  $\xi$ and $\Delta p$, as  originally proposed by Gondek-Rosińska {\it et al.}~\cite{Gondek:1997fd}, which {in the charged case} are given by (assuming a harmonic dependence on the unknown functions)
\begin{widetext}
  \begin{align}
    \label{eq:Gondek1}
    \xi' & = -\frac{1}{r}\left(3\xi + \frac{\Delta p}{\gamma p_0}\right) + \nu_0'\xi\,,\\
    \label{eq:Gondek2}   
    (\Delta p)' & = \xi\left[r(\epsilon_0 + p_0)\left(\omega^2 e^{2(\lambda_0 - \nu_0)} + \nu_0'^2 - 8\pi e^{2\lambda}p_0 - \frac{q_0^2}{r^4}\right) - 4p_0'\right] -\Delta p\left[\nu_0' + 4\pi r(\epsilon_0 + p_0)e^{2\lambda_0}\right] \,,
  \end{align}
\end{widetext}
where $\gamma$ is the adiabatic index, $(')$ denotes a total radial derivative and the subindex 0 represents the quantities calculated in the background (unperturbed) metric. Such equations can only be solved after finding the thermodynamic profiles through the TOV equations for a given central energy density (pressure). For more details on the TOV equations accounting for the presence of electric charge in the star, see~\cite{Goncalves:2020joq} and references therein. Finally, in order to determine unique solutions, one imposes the following boundary conditions: $(\Delta p)_{r=0} = -3(\xi\Gamma p)_{r=0}$ at the stellar center and $(\Delta p)_{r=R} = 0$ on the surface. One also assumes that $\xi$ is normalized at the center, i.e. $\xi(0) = 1$. {Notice that the neutral limit is reached when $q_{0}\to 0$ explicitly in Eqs. (\ref{eq:Gondek1})--(\ref{eq:Gondek2}) and implicitly in the metric functions, i.e. $\nu_{0}(q_{0}\to 0)$ and $\lambda_{0}(q_{0}\to 0)$.}

In the last decades, several detailed calculations on the stability of one-phase {neutral} stars (see, e.g., Refs.~\cite{Glendenning:2000,Lugones:2021bkm}) proved that when the EoS is smooth enough for all the available energy densities, i.e. a {sharp} first-order phase transitions such as the confinement/deconfinement transition is not present, the stability configurations can be determined by  the analysis of the criterion $\partial{M}/\partial p_{c} \geq 0$
 instead of solving the aforementioned coupled set of equations.
 Nevertheless, recent studies as e.g. Ref.~\cite{Pereira:2017rmp}, have demonstrated  that a sharp discontinuity in the EoS non-trivially affects the dynamical stability of the hybrid NS whether the microphysical phase-conversion dynamics happens faster or slower than the dynamical timescale of the fundamental radial oscillation mode, $\tau^{n=0}_{\text{dyn}}$. More precisely, slow conversions could still satisfy $\omega^{2}_{n=0} \geq 0$ even when $\partial{M}/\partial p_{c} < 0$, whereas rapid conversions always satisfy $\partial{M}/\partial p_{c} \geq 0$. Thus, for the former case one is immediately led to consider the possibility of a new branch of stable hybrid stars,  the so-called twin and triplet stars\footnote{Notice that ordinary triplet stars have been studied e.g. in Refs. \cite{Alford:2017qgh,Han:2018mtj,Li:2019fqe,Li:2021sxb}  where the assumption of slow conversions was not applied when analyzing their
stability. In our notation, ordinary triplet stars mean triplet stars with rapid transitions where the criteria of $\partial M / \partial p_c > 0$ still works.}, as already pointed out in Ref.~\cite{Lugones:2021bkm}.  Such results indicate that  in order to determine the properties and stability window of twin stars, we should solve the Gondek-Rosińska {\it et al.} equations considering the junction conditions associated to slow and rapid conversions.

 Mathematically, one implements these junction conditions at the interface between different phases~\cite{Pereira:2017rmp} considering the phase-transition speed. One has that for both conversion speeds, the boundary is kept in thermodynamic equilibrium during the radial oscillations which in turn implies $\Delta p^+ - \Delta p^- = 0$, where + (-) represents the Lagrangian perturbation of the pressure after (before) the phase transition. {The difference between conversion speeds implies distinct extra boundary conditions for the relative radial displacement \cite{Haensel:1989wax}. For slow transitions, volume elements near the sharp boundary do not change their nature but instead co-move with it, which implies that $\xi$ is continuous in a phase splitting surface. On the other hand, rapid conversions of elements near the surface imply that the relative radial displacement is no longer continuous~\cite{Pereira:2017rmp}} then producing the following additional condition:
\begin{equation}
\xi^+ - \xi^- = \frac{\Delta p}{r}\left(\frac{1}{p_{0}^{'+}} - \frac{1}{p^{'-}_{0}}\right) \,,
\end{equation}
with  $p'_0 \equiv p'_0(r)$  being the derivative of the unperturbed  pressure.  These extra boundary conditions can be formally derived using general distributions according to the formalism of Ref.~\cite{Pereira:2015sua}, where they have emphasized that the extra boundary conditions do not change when one considers the presence of electric charge. In this sense, the changes are contained in the unperturbed metric functions $\nu_0$ and $\lambda_0$, alongside the extra term in Eq.~(\ref{eq:Gondek2}). 

In order to solve the Gondek-Rosińska {\it et al.} and TOV equations for {neutral and electrically charged} twin stars we should specify the EoS for hadronic and quark matter. As in Ref.~\cite{Goncalves:2021pmr}, we will consider two types of EoSs for quark matter: the constant speed of sound (CSS) parametrization~\cite{Alford:2004pf,Alford:2013aca}, {and the multipolytropes (MP) approach  proposed in Ref.~\cite{Alvarez-Castillo:2017qki}} (see Ref.~\cite{Goncalves:2021pmr} for further details). One has that in the CSS parametrization, the speed of sound is taken to be constant for the QM phase, which is an excellent approximation at high densities~\cite{Alford:2004pf,Alford:2013aca}. In our calculations, we will assume that the speed of sound is equal to the speed of light in order to have the stiffest possible QM EoS. In contrast, in the MP approach, the speed of sound is assumed to be dependent on the density. On the other hand, the hadronic phase will be described assuming that the EoS is given by a generalized piecewise polytropic (GPP)~\cite{OBoyle:2020qvf} with parameters given by $\log_{10}K_1 = -27.22$, $\Gamma_{1,2,3} = (2.764, 7.75, 3.25)$ which matches the chiral-effective stiffest EoS of Hebeler {\it et al.}~\cite{Tews:2012fj,Hebeler:2013nza} at $1.1{n_0}$. For the crust, i.e. up to 0.3 $n_{0}$, we used the SLy4 EoS (Table II of Ref.~\cite{OBoyle:2020qvf}). {In this work, we will assume this single hadronic EoS for all QM models, which means that all configurations will share the same hadronic branch for central pressures up to the transitional pressure.} Finally, for the case of charged twin stars, we will assume that the charge distribution is proportional to the energy density, i.e., $\rho_e = \alpha\epsilon\,,$ where, in geometric units, $\alpha$ is a dimensionless proportionality constant which can be considered a free parameter in our modeling. Such distribution is usually denoted by $\alpha$-distribution and have been considered in previous studies related to the stability of charged compact stars~\cite{Brillante:2014lwa,arbanil2015,Panotopoulos:2019wsy,Goncalves:2020joq,Panotopoulos:2020hkb,Jasim:2021cft,Goncalves:2021pmr}. It is important to emphasize that $\alpha$ should be smaller than one in order to avoid the collapse of the compact star into a black hole.

\section{Results}
\label{sec:results}

\begin{figure*}[!t]
  \centering
  \includegraphics[width=.45\textwidth]{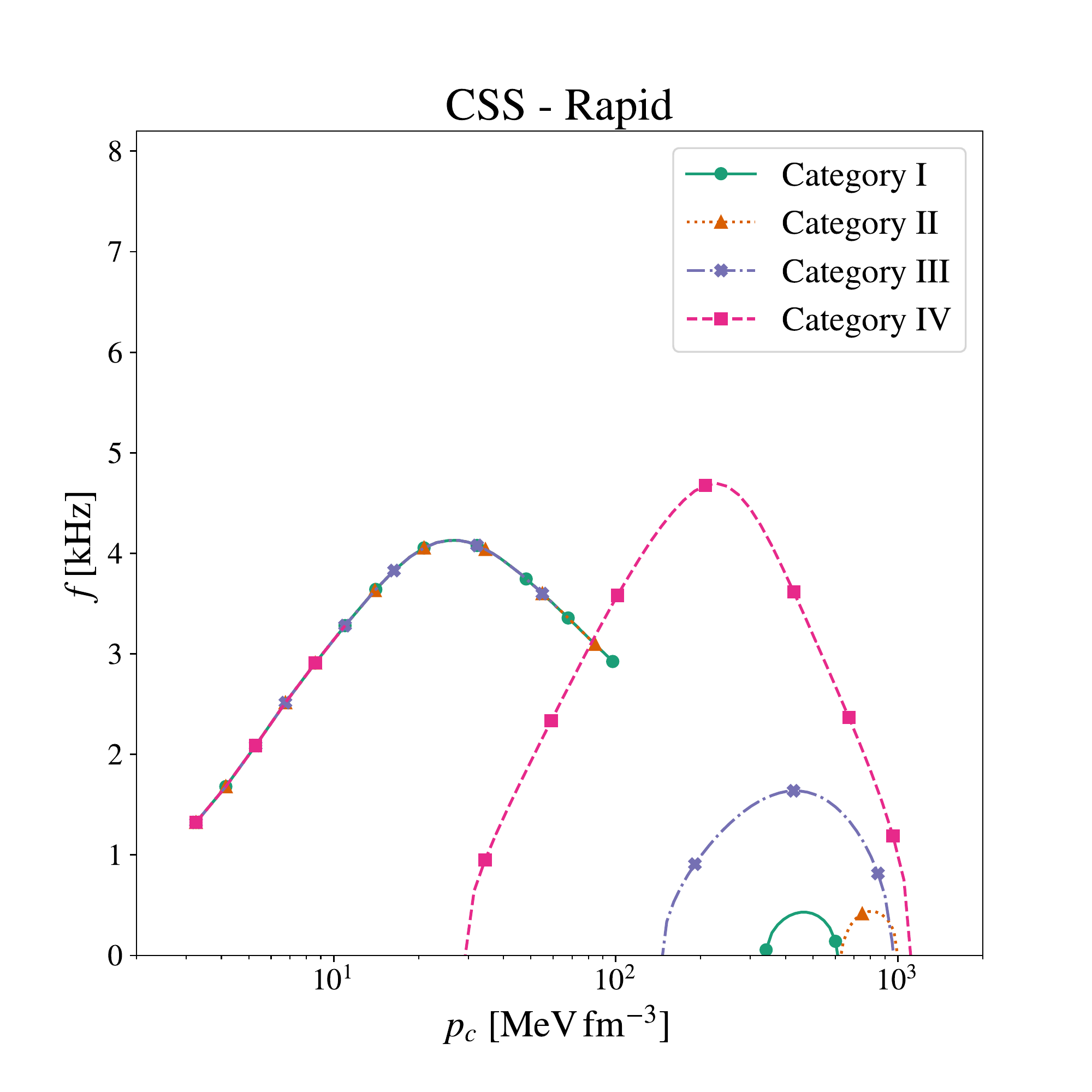}%
  \hfill%
  \includegraphics[width=.45\textwidth]{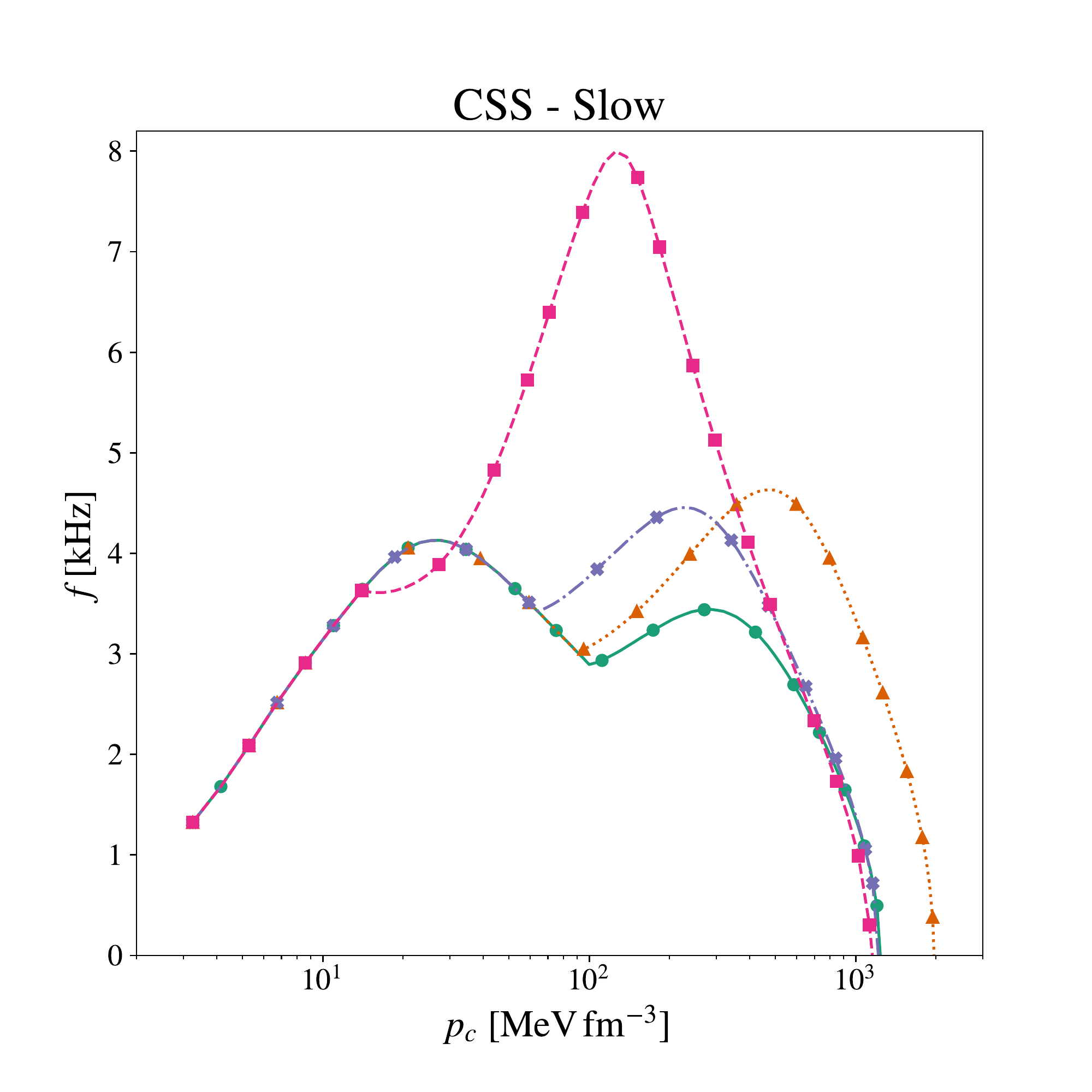}
  \caption{Predictions  for the fundamental linear eigenfrequency as a function of central pressure for neutral twin stars, considering rapid (left panel) and slow (right panel) conversions, considering the Categories I to IV of the CSS parametrization for the QM EoS.}
  \label{fig:fxP_neutral-1}
\end{figure*}

In what follows, we will investigate the impacts of highly discontinuous transitions on the dynamical stability of twin hydrid NSs, in particular, those obtained in our past work~\cite{Goncalves:2021pmr}. As explained in the previous section, we will assume the CSS parametrization and the MP approaches for QM and a single hadronic EoS, which means that all configurations will share the same hadronic branch for central pressures up to the transitional pressure. Besides, the QM parameters were chosen in order to satisfy the Seidov constraint~\cite{Seidov:1971}, which generates, using the usual stability criterion $\partial M/\partial p_c \geq 0$, hybrid stars in a disconnected branch from the hadronic one, the so-called third family of NSs. In this usual sense, twin stars arise from such a discontinuity. However, here we are simply considering that twin stars are stars that have the same mass but different radii.

Initially, let us present our results derived considering the CSS parametrization for the QM EoS. Following Ref.~\cite{Christian:2020xwz}, one can distinguish four categories of twin stars in this parametrization: (a) Category I: both stars have maximum masses larger than 2 $M_\odot$; (b) Category II: only hadronic stars reach masses larger than 2 $M_\odot$; (c) Category III: the hadronic stars have a maximum mass between 1 and 2 solar masses, while the hybrid one surpasses 2 $M_\odot$; (d) Category IV: the hadronic stars maximum mass is smaller than 1 $M_\odot$, while the hybrid one is larger than 2 $M_\odot$. The associated predictions for the fundamental mode eigenfrequencies, derived assuming a neutral hybrid star ($\alpha = 0$) and a rapid (slow) conversion, are presented in the left (right) panel of Fig.~\ref{fig:fxP_neutral-1}. {Note that, for convenience, we are plotting the positive real part of the linear frequency which is simply $f=\omega/(2\pi)$, when $\omega_{n=0}^2 \geq 0$.} Our calculations show that twin stars in Categories I to IV display quantitative and qualitative differences regarding the behavior of the zero-mode linear frequencies when assuming a rapid or slow conversion process. One can see that dependending on the transition speed one finds disconnected (rapid) or connected (slow) branches of ultra-dense electrically neutral stellar configurations. As one can also see in this figure, all categories share the same hadronic branch, in agreement with the results presented in the right panel of Fig.~1 in Ref.~\cite{Goncalves:2021pmr}. In addition, the magnitudes of the frequencies for hybrid stars are larger for slow conversions when compared to rapid ones, as expected from the results obtained in Ref.~\cite{Pereira:2017rmp}. 

One also has that in the case of rapid conversions, twin star configurations associated to Category 4 behave quite differently in comparison to the other categories, with the maximum frequency for the hadronic branch being larger than the quark one, i.e.  max($f_{\rm had}$) $<$ max($f_{\rm twin}$). In constrast, the remaining categories present max($f_{\rm had}$) $>$ max($f_{\rm twin}$). Notice that in Categories I and II the max ($f_{\rm twin}$) are smaller than any of the other results, which could even be compared with modifed theories of gravity~\cite{Mendes:2018qwo}. On the other hand, for slow conversions, the only category that does not satisfy max($f_{\rm had}$) $<$ max($f_{\rm twin}$) are Category I configurations. In particular, when slow conversions take place, even triplet stars i.e. three stellar configurations that have the same mass but different radii, may occur for a single sharp phase transition with connected branches. 

\begin{figure*}[t!]
  \centering
  \includegraphics[width=.45\textwidth]{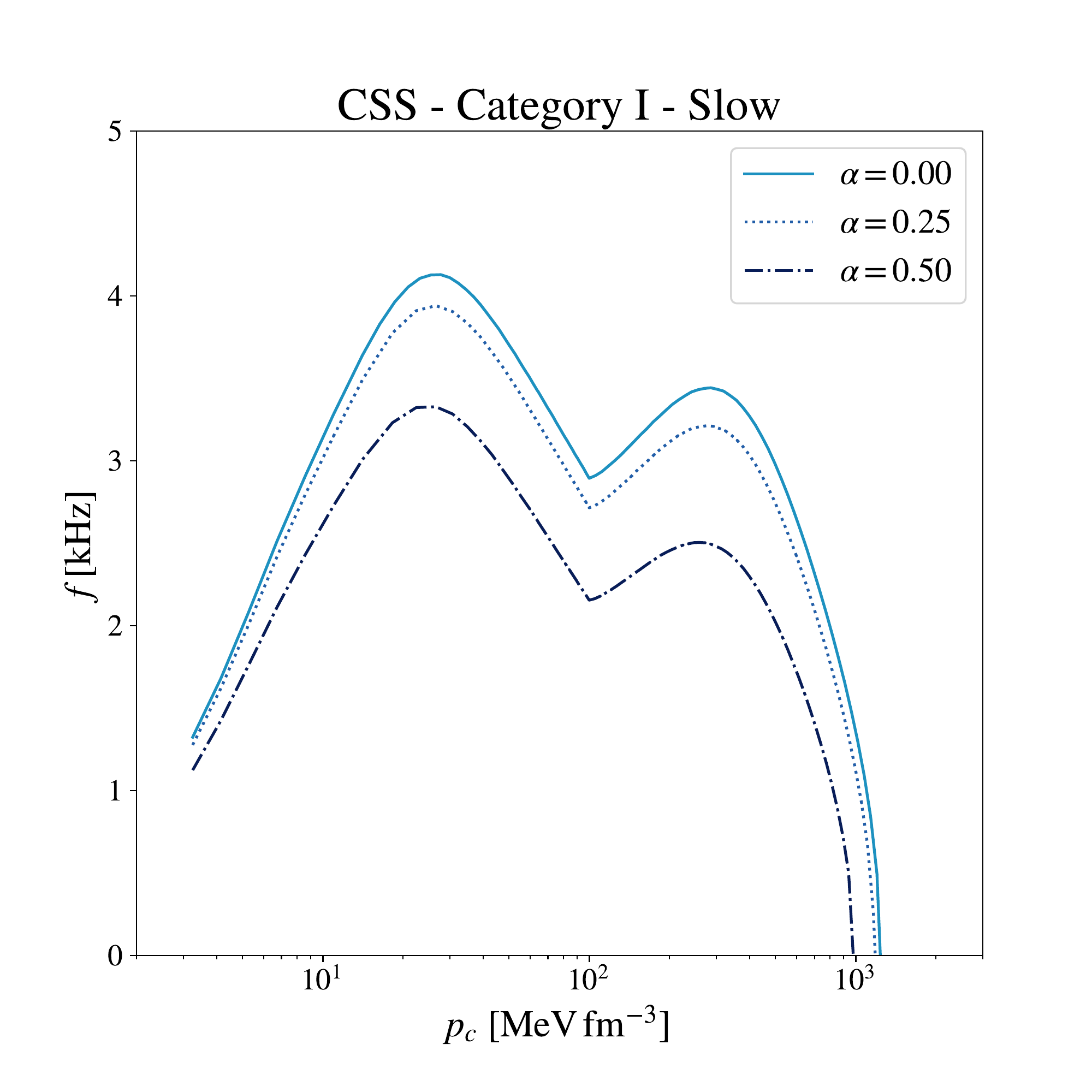}%
  \hfill%
  \includegraphics[width=.45\textwidth]{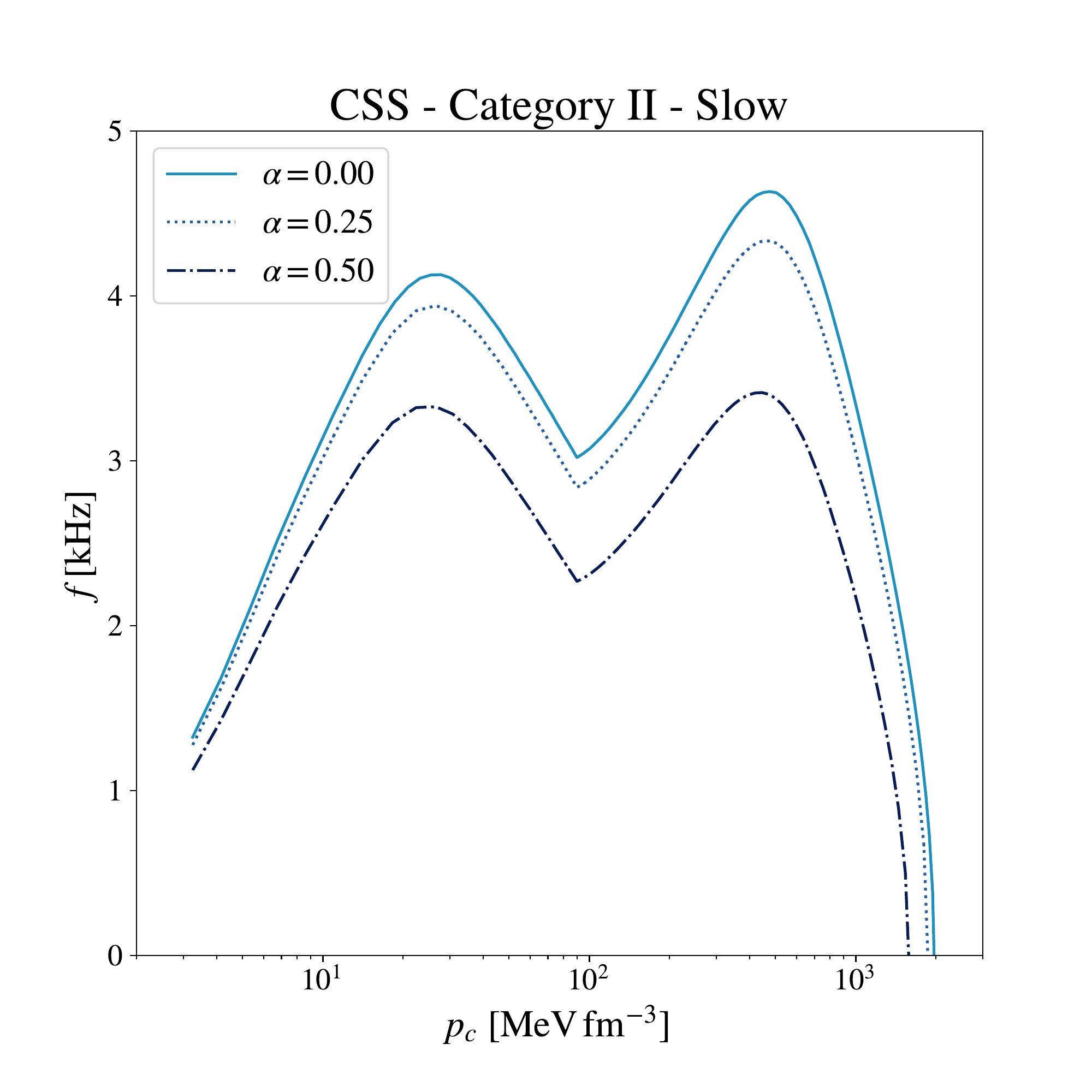}
  \caption{Predictions for the fundamental linear eigenfrequency as a function of central pressure for charged twin stars considering slow conversions and the Categories I (left panel) and II (right panel) of the CSS parametrization. {Note that the frequencies are not in the same scale as Fig.~\ref{fig:fxP_neutral-1}.}}
  \label{fig:fxP_cat12-2}
\end{figure*}

\begin{figure*}[!t]
  \centering
  \includegraphics[width=.45\textwidth]{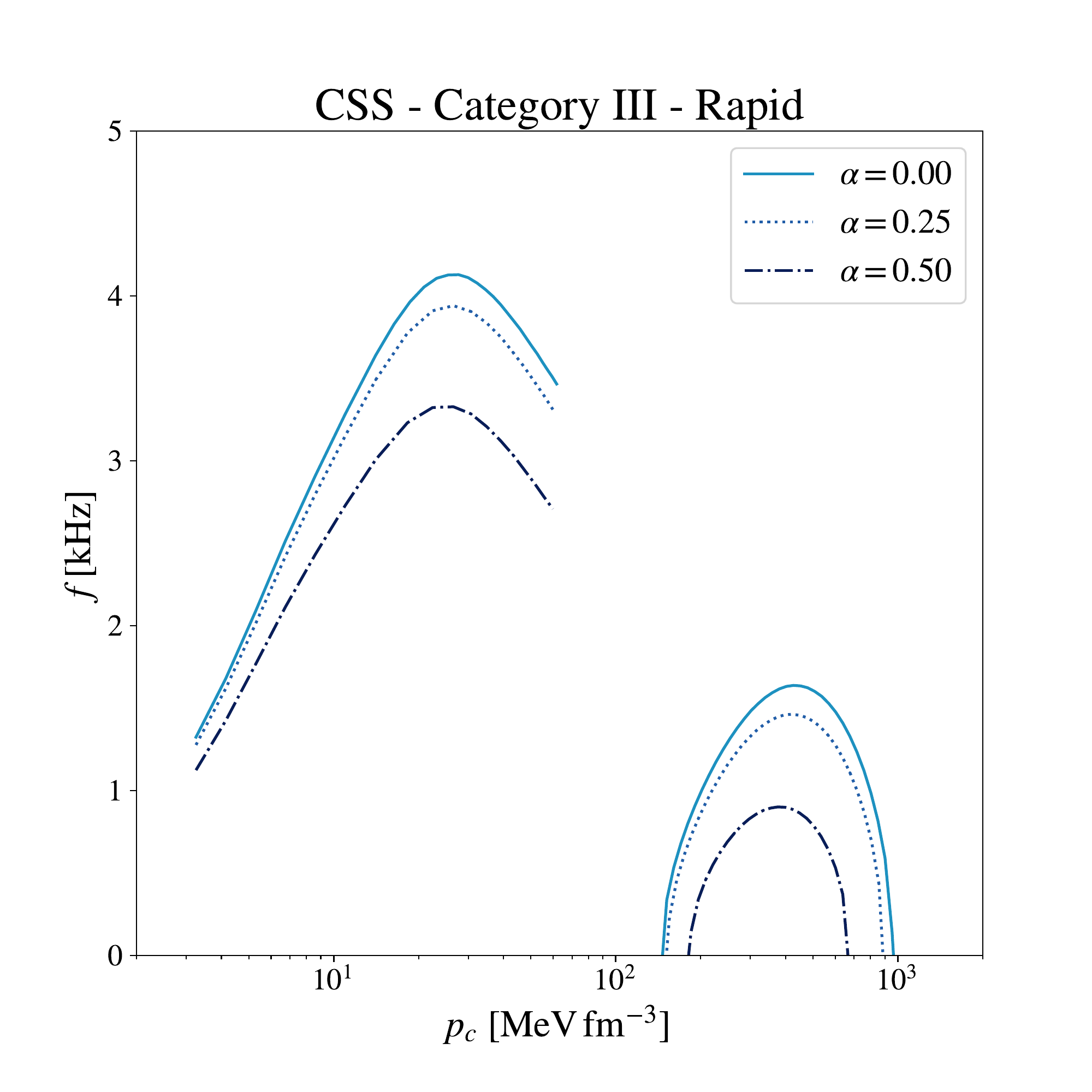}%
  \hfill%
  \includegraphics[width=.45\textwidth]{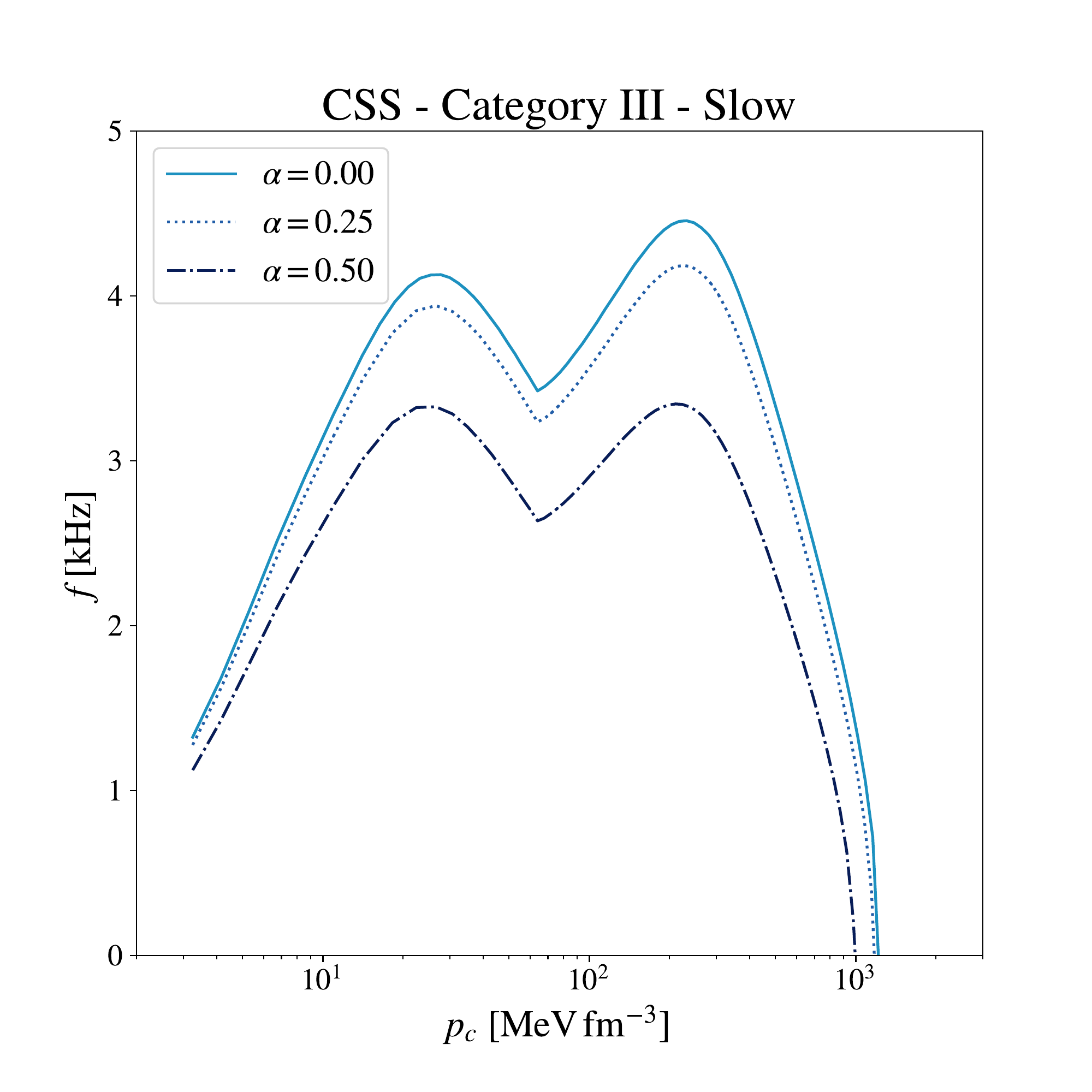}
  
  \includegraphics[width=.45\textwidth]{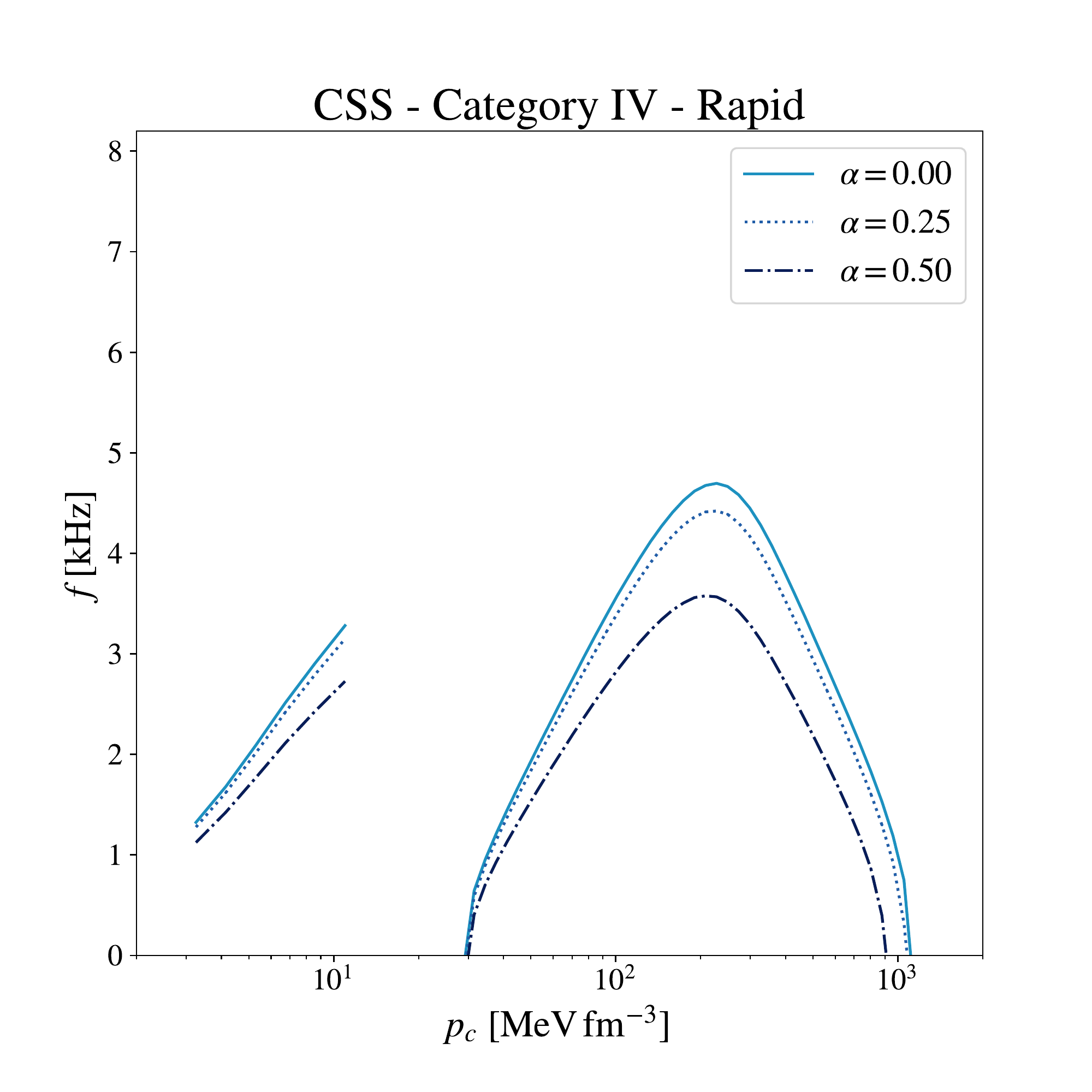}%
  \hfill%
  \includegraphics[width=.45\textwidth]{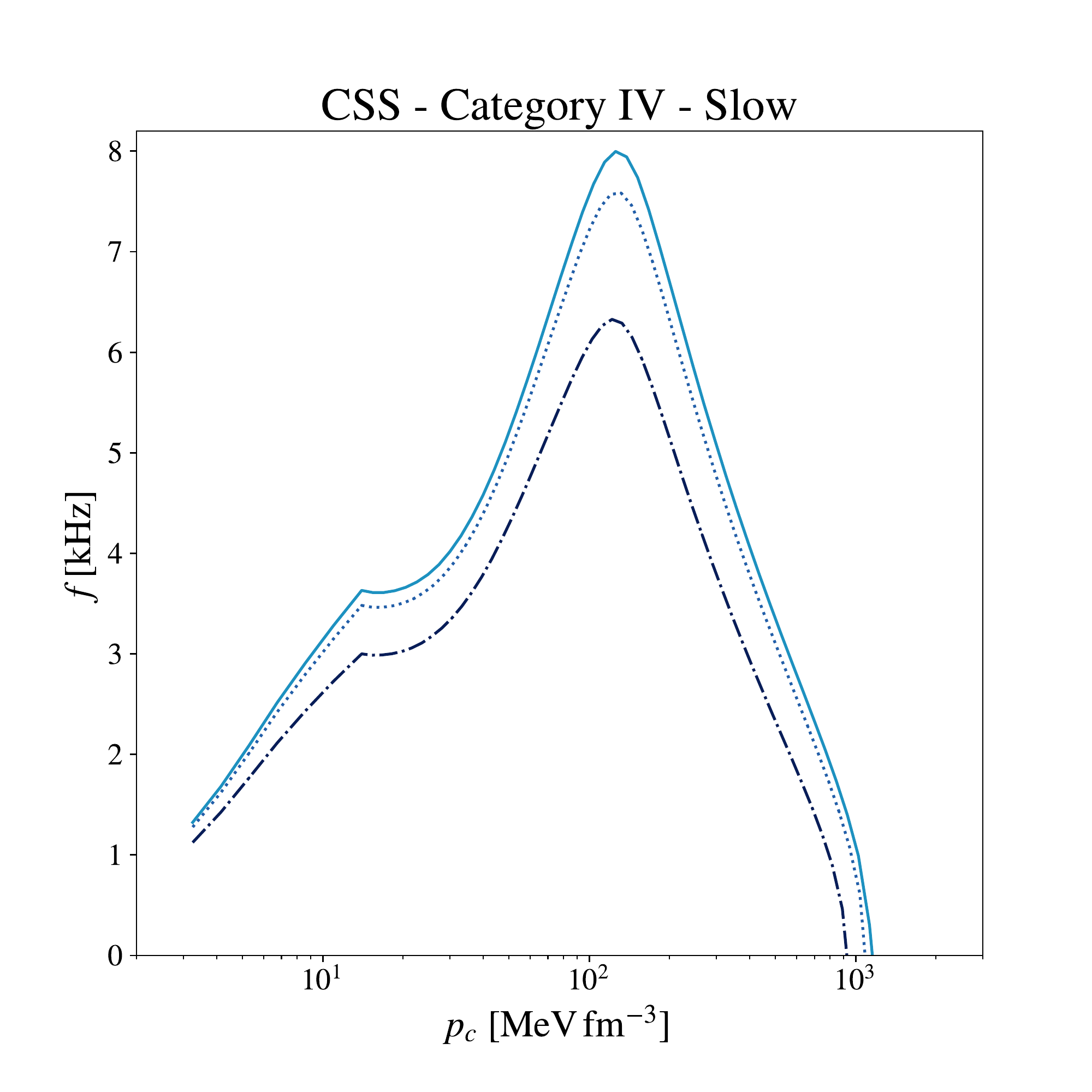}
  \caption{Predictions  for the fundamental linear eigenfrequency as a function of central pressure for charged twin stars considering rapid (left panel) and slow (right panels) conversions and the Categories III (upper panels) and IV (lower panels) of the CSS parametrization. {Note that the frequencies are not in the same scale for the different categories.}}
  \label{fig:fxP_cat3-3}
\end{figure*}

 The impact of electric charge on the stability of twin hybrid NSs described by the CSS parametrization and assuming the $\alpha$-distribution is analyzed in Figs.~\ref{fig:fxP_cat12-2} and \ref{fig:fxP_cat3-3}, where we present the predictions for $\alpha=0.25$ and $0.5$. For comparison, the results for the neutral case ($\alpha = 0$) are also presented. When rapid conversions are assumed, one has that Categories I and II do not display any stable configuration under radial perturbations. This can be understood by looking at the correspoding mass-radius diagram where only a small set of configurations survive under the criterion $\partial M/ \partial p_c \geq 0$ but which when subjected to the dynamical-stability analysis prove to be unstable (see Ref.~\cite{Goncalves:2020joq} for similar results in other compact stars). As a consequence, for these categories, the stable configurations only arise for slow conversions. The associated predictions for the fundamental mode eigenfrequencies are presented in Fig.~\ref{fig:fxP_cat12-2}. One has that the main impact is on the normalization, with the shape being similar, even in the case of $\alpha=0.5$, which is by no means a trivial amount of electric charge. As in the neutral case, the hadronic and ultra-dense branches are still connnected.

For charged twin stars associated to the Categories III and IV, one has verified that stable configurations are present for slow and rapid conversions. The results are presented in Fig. \ref{fig:fxP_cat3-3}.  Generically, one can see that the increasing of the amount of electric charge (higher $\alpha$) diminishes the magnitude of the frequencies and the stability window. {This is in agreement with the results presented in our previous work~\cite{Goncalves:2021pmr}, where we have shown that increasing $\alpha$ reduced the region where $\partial M/ \partial p_c < 0$. In this sense, for rapid conversions some charged configurations that satisfy the usual stability criterion are dynamically unstable, reducing even further the stability window for charged twin stars.}

\begin{figure*}[!t]
  \centering
  \includegraphics[width=.45\textwidth]{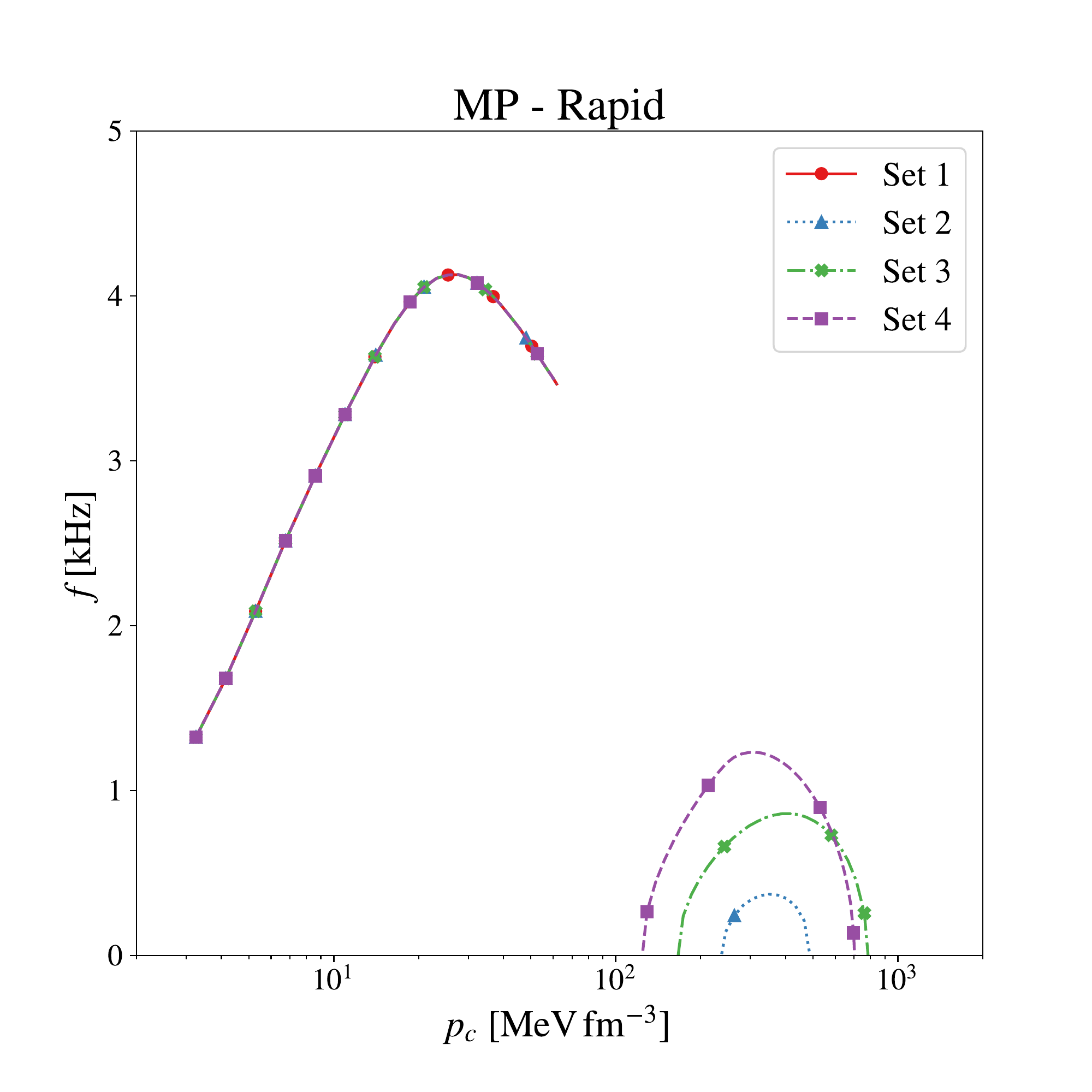}%
  \hfill%
  \includegraphics[width=.45\textwidth]{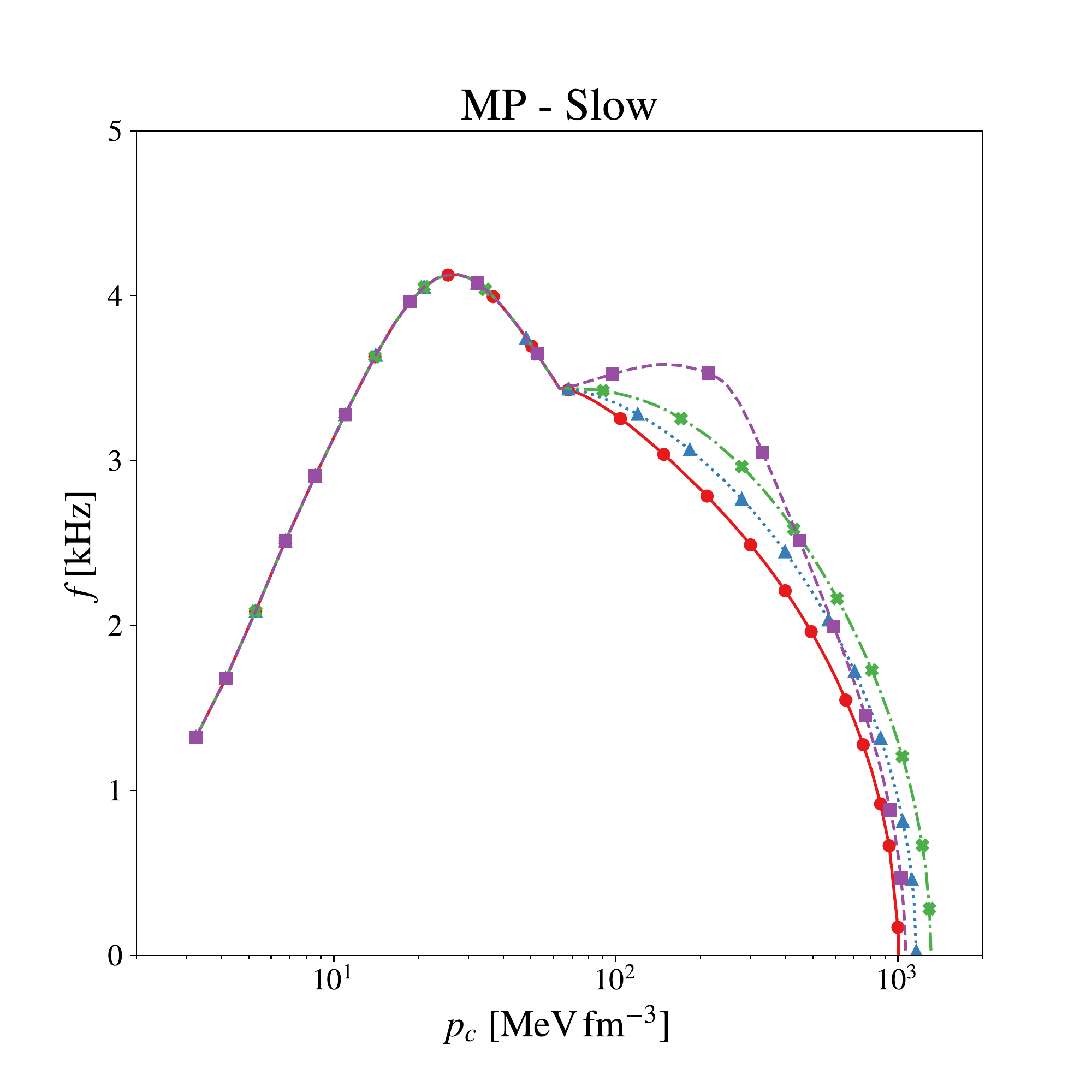}
  \caption{Predictions  for the fundamental linear eigenfrequency as a function of central pressure for neutral twin stars, considering rapid (left panel) and slow (right panel) conversions and the distinct sets of the MP approach for the QM EoS.}
  \label{fig:fxP_MP_neutral-4}
\end{figure*}

Let us now consider that the EoS for quark matter is described by the MP approach proposed in Ref.~\cite{Alvarez-Castillo:2017qki}, in which the speed of sound is dependent on the density. We consider this distinct approach for the quark matter EoS in order to test the robustness of our findings and extract generic conclusions about the properties of twin hybrid stars. In Ref.~\cite{Alvarez-Castillo:2017qki}, the authors have proposed four different sets of EoS, with each set presenting an EoS that is stiffer than the previous one. In particular, Set 4 is described by two polytropes in order to not violate causality. In the aforementioned category scheme, all four sets generate Category III twin stars. For additional details, we refer the reader to our previous work~\cite{Goncalves:2021pmr}.

In Fig. \ref{fig:fxP_MP_neutral-4}, we present our predictions for the fundamental eigenfrequency considering a neutral twin star and the different sets of the MP approach. The results for a rapid (slow) conversion are presented in the left (right) panel. For a rapid conversion, only Set 1 does not generate a stable configuration at high densities. For the other sets, we predict two disconnected branches. In contrast, for a slow conversion, one finds that all sets predict stable configurations in connected branches. It should be stressed that quantitatively, in the rapid case, the hadronic and twin branches have considerable differences, i.e. the hadronic branch is almost {4 times} larger than the twin branch which could be useful in future measurements of twin star properties to distinguish faithfully between both kinds of compact stars or even compare with similar results within modified theories of gravity~\cite{Mendes:2018qwo}.

\begin{figure*}[!t]
  \centering
  \includegraphics[width=.45\textwidth]{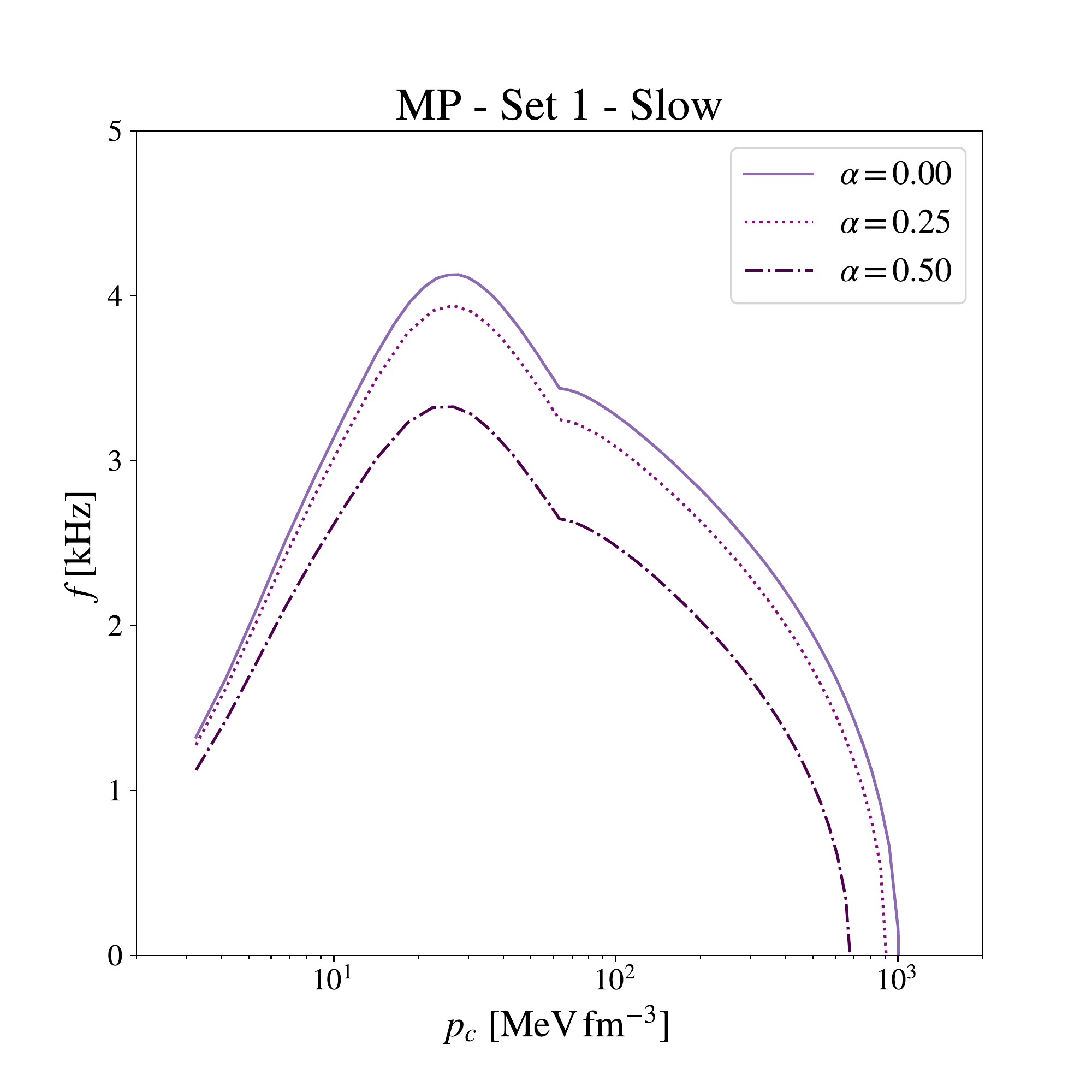}%
  \hfill%
  \includegraphics[width=.45\textwidth]{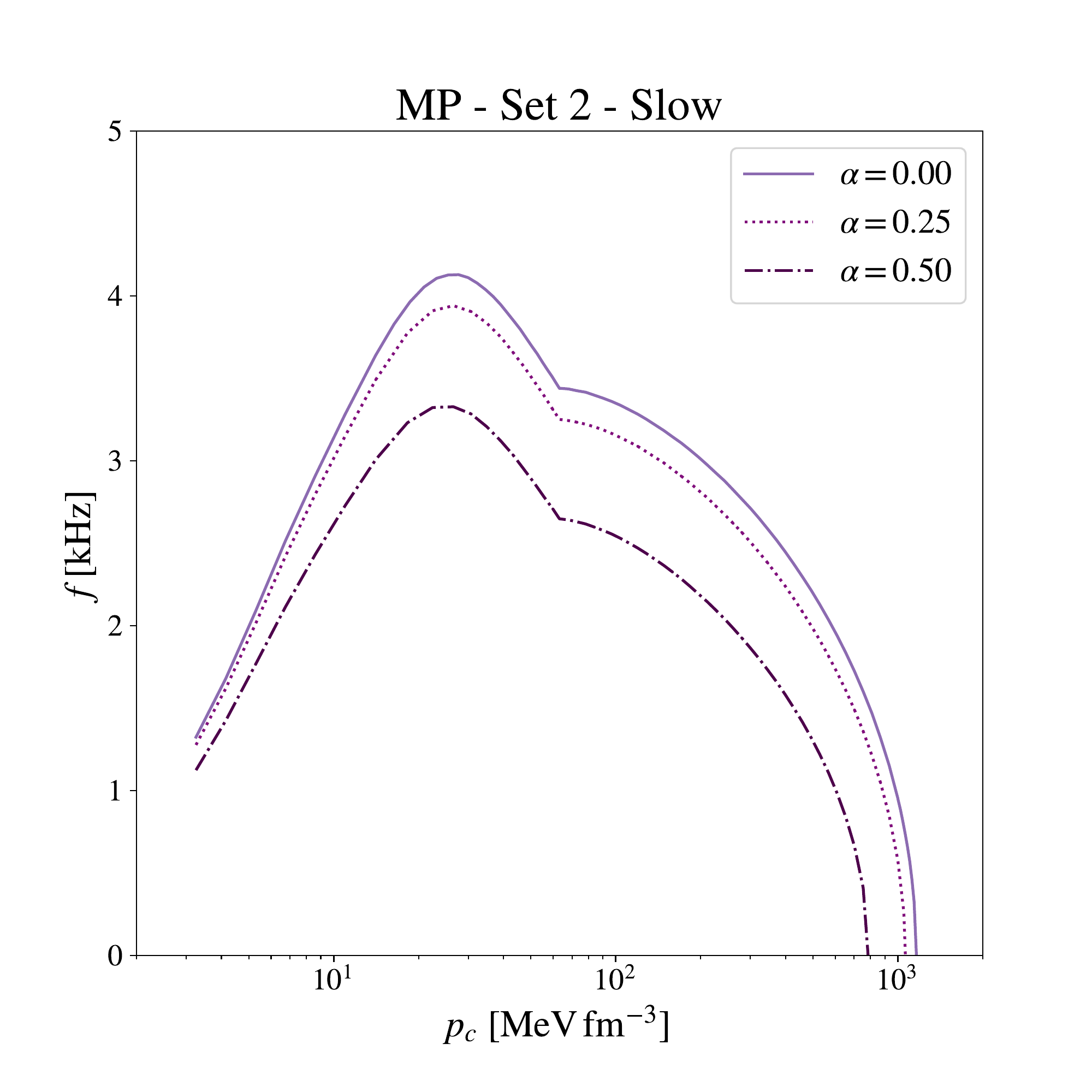}
  \caption{Predictions  for the fundamental linear eigenfrequency as a function of central pressure for charged twin stars, considering slow conversions and the Set 1 (left panel) and Set 2 (right panel) of the MP approach for the QM EoS.}
  \label{fig:fxP_MP_set12-5}
\end{figure*}

The predictions for charged twin stars, derived using the MP approach for the QM EoS, are presented in Figs.~\ref{fig:fxP_MP_set12-5} and~\ref{fig:fxP_MP_set34-6}. One has verified that Sets 1 and 2 do not present stable configurations in the case of rapid conversions. However, under slow conversions they present configurations with real fundamental eigenfrequencies that can be called slow-stable. The corresponding results are presented in Fig.~\ref{fig:fxP_MP_set12-5}. Similarly to what was verified in Fig.~\ref{fig:fxP_cat12-2} using the CSS parametrization, one has that the presence of charge decreases the magnitude of the eigenfrequency and diminishes the stability window. In contrast, for $\alpha = 0.25$ and the Sets 3 and 4, which are stiffer than the previous sets, one has verified that stable configurations are present for slow and rapid conversions, for which the predictions are shown in Fig.~\ref{fig:fxP_MP_set34-6}. On the other hand, for $\alpha = 0.5$, stable configurations are not present in the high density regime for Set 3 under rapid conversions. For all these sets, one has that ultra-dense twin stars have very low frequencies when compared to the hadronic branch and then are immediately distinguishable from nuclear-matter stars. From these results, one can clearly see that the stellar configurations from the MP sets are very similar to Category III twin stars from the CSS parametrization. In fact, the distinct behaviors of both QM models regarding the speed of sound does not change the qualitative behavior of the fundamental linear frequency.
          
\begin{figure*}[!t]
  \centering
  \includegraphics[width=.45\textwidth]{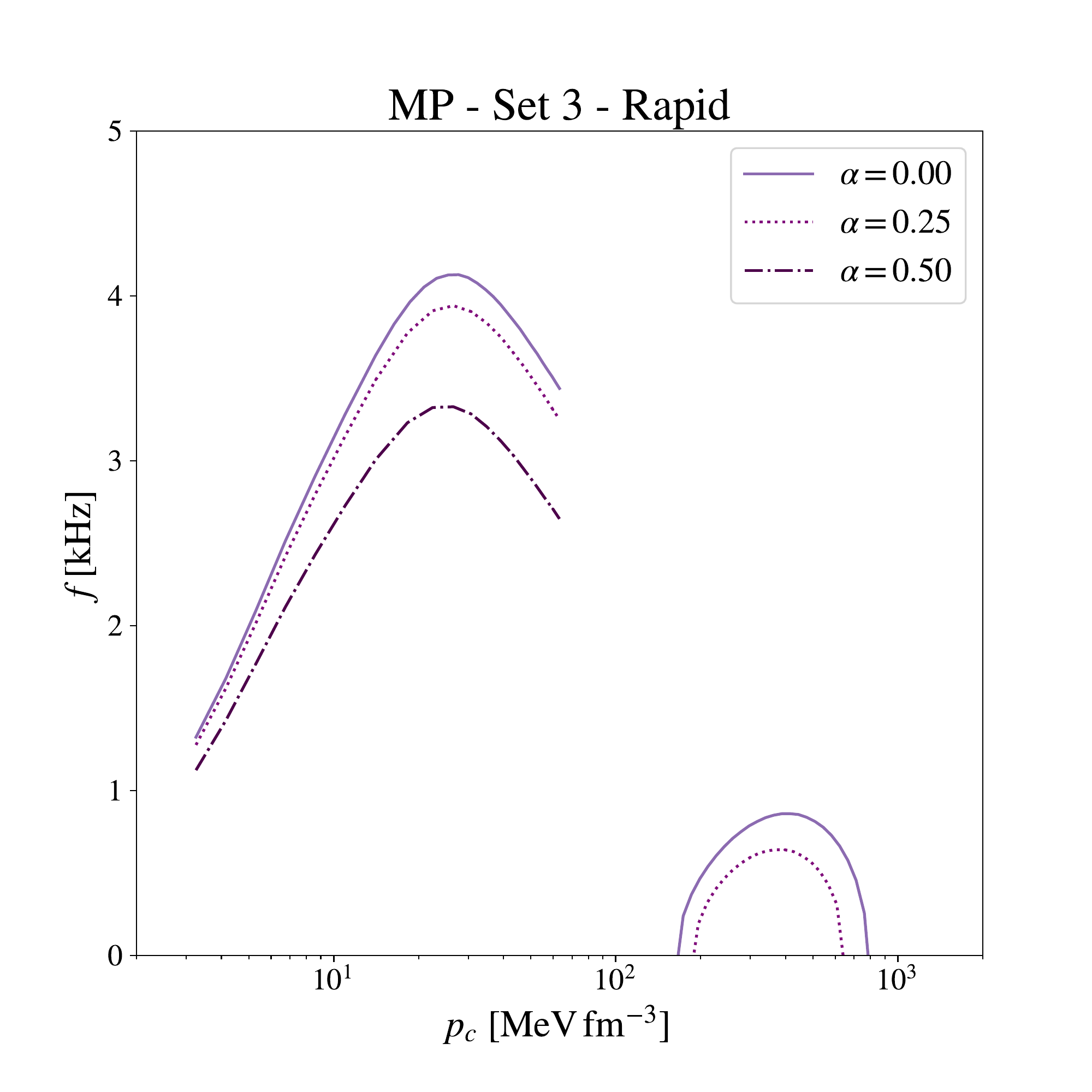}%
  \hfill%
  \includegraphics[width=.45\textwidth]{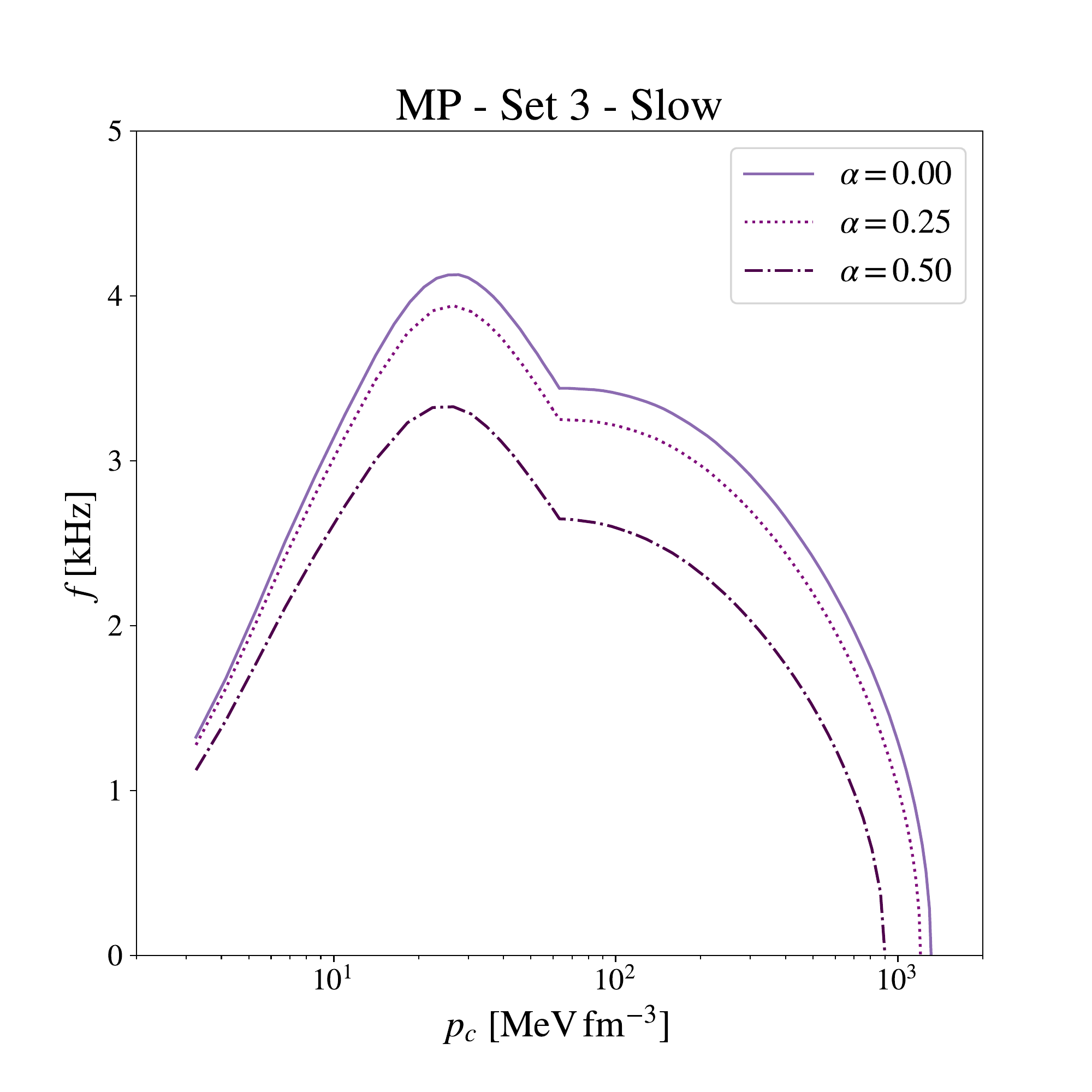}
  
  \includegraphics[width=.45\textwidth]{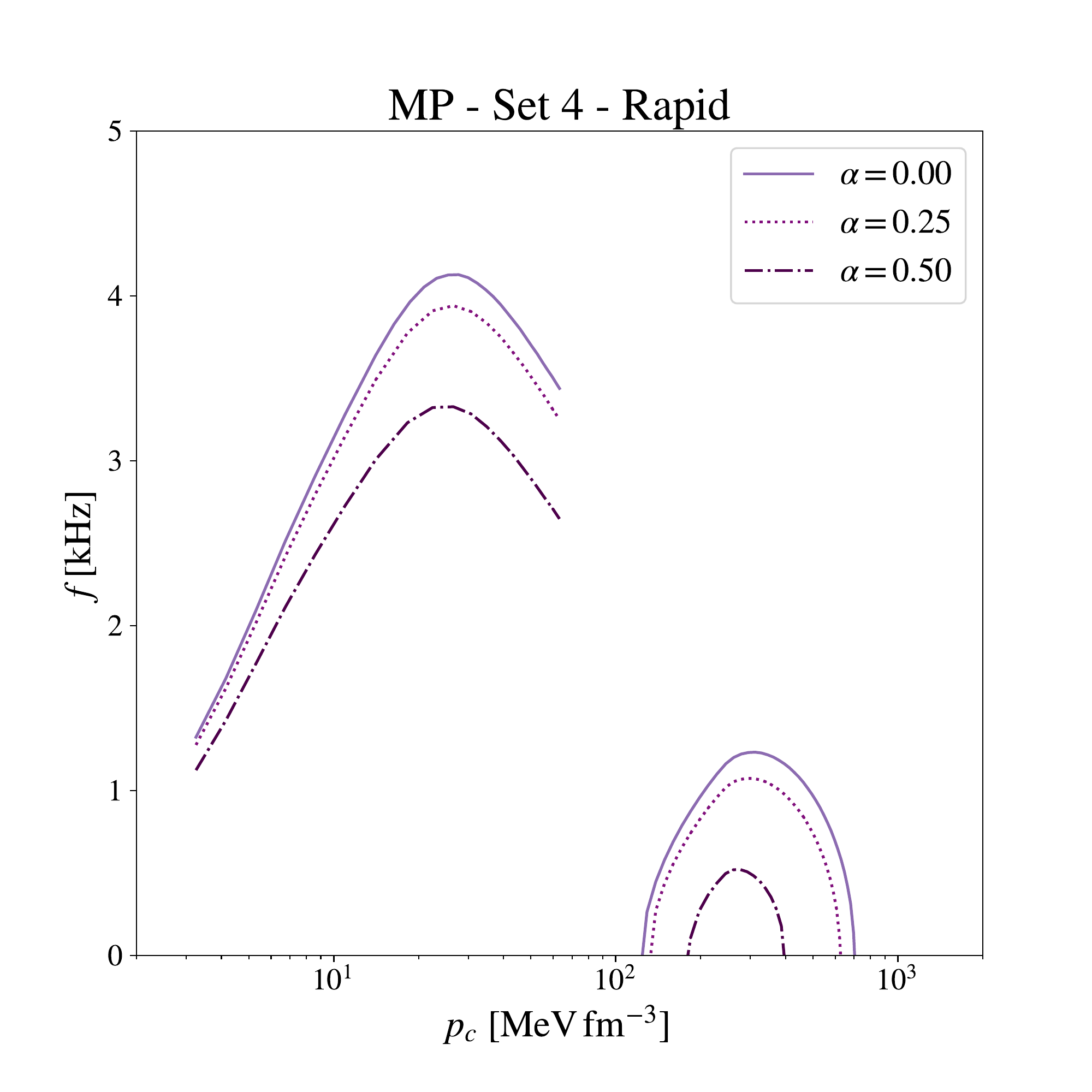}%
  \hfill%
  \includegraphics[width=.45\textwidth]{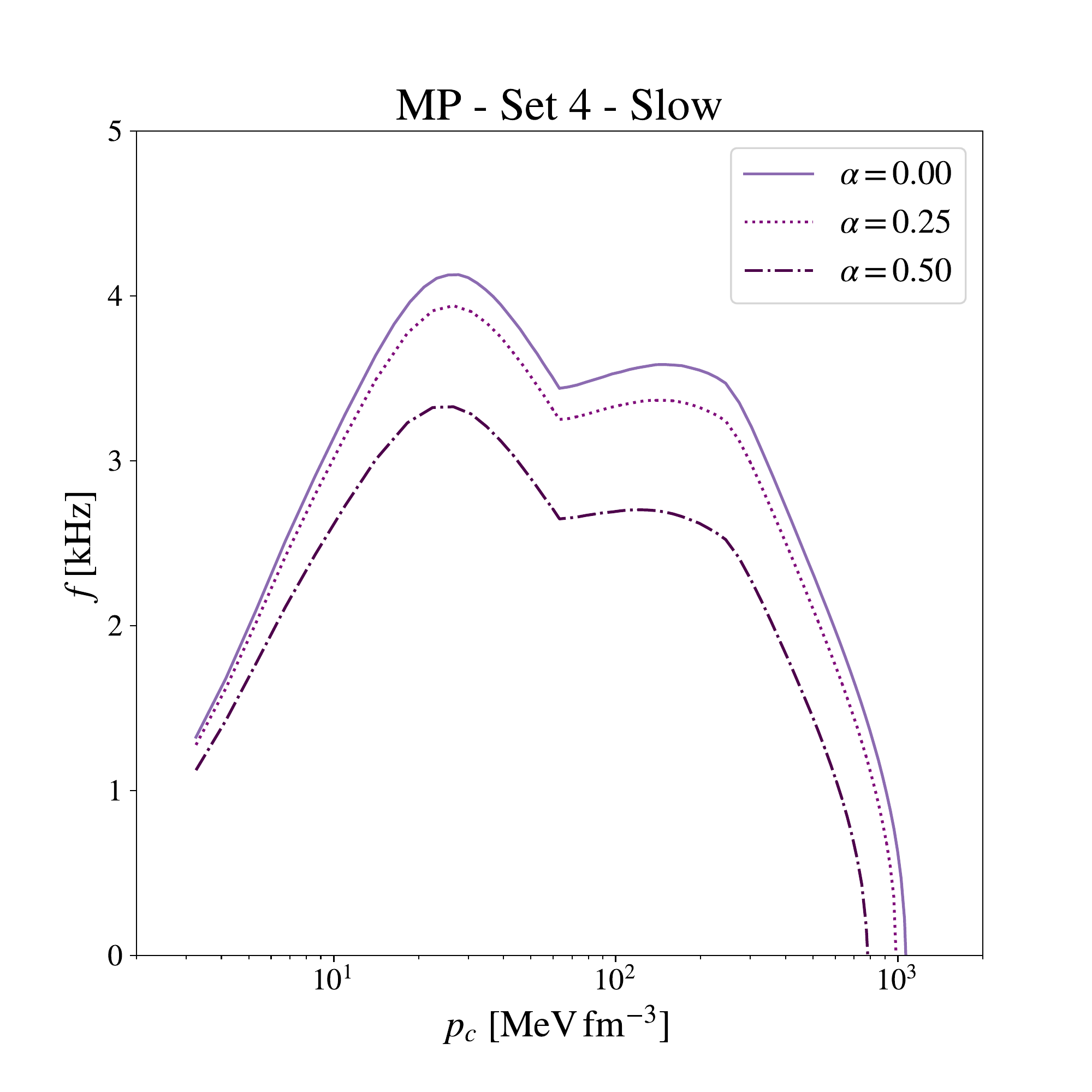}
  \caption{Predictions  for the fundamental linear eigenfrequency as a function of central pressure for charged twin stars, considering rapid (left panels) and slow (right panels) conversions and the Set 3 (upper panels) and Set 4 (lower panels) of the MP approach for the QM EoS.}
  \label{fig:fxP_MP_set34-6}
\end{figure*}

\section{Conclusions}
  \label{sec:conclusion}

In this work, we have investigated systematically the effects of rapid and slow phase conversions on the dynamical stability of neutral and electrically charged twin hybrid neutron stars when strong first-order transitions occur between hadronic and quark matter at their cores. For low and intermediate densities, GPP parametrizations and results from chiral-effective field theory were used respectively when obtaining the associated equations of state. On the other hand, the ultra-dense EoSs for quark matter were modeled independently within the CSS and MP parametrizations in order to test the robustness of our results.

{We did this by performing} a radial pulsation analysis focusing on the fundamental-mode {eigenfrequencies} in order to probe sizeable qualititative and quantitative differences compared to one-phase and two-phase stars with smooth transitions at twin-star interiors \cite{Brillante:2014lwa}. In order to explore a larger parameter space, we also considered finite distributions of electric charge with conservative values of the free parameter $\alpha$. {In turn, this led us to use carefully the charged pulsation formalism when large jumps of energy density are proposed for the first-order transition.} 

In the neutral case, our calculations indicate that slow conversions connect initially disconnected branches thus precluding the existence of a third family of NSs, in the usual sense, but at the same time favoring the twin star possibility even leading to triplet stars, i.e. three stellar configurations having the same mass. Also, twin-star configurations in all categories of the CSS and MP equations of state for rapid conversions show that their radial oscillation linear frequencies are considerably smaller, being immediately distinguishable in future non-radial oscillation measurements, which are non-linearly coupled to the radial ones in gravitational-wave data.

On the other hand, in the  {(still hypothetical)} electrically charged case, we have found similar conclusions for which in some cases of the parametrizations only quantitative increments are found with no sizeable qualitative differences. Moreover, we have found that the presence of electric charge diminishes the stability window of twin stars, but increases their masses and radii, in agreement with our previous work~\cite{Goncalves:2021pmr}. As in the case of one-phase charged stars~\cite{Goncalves:2020joq}, it also presents a scenario where the usual stability criterion is no longer sufficient even when rapid conversions take place.

Our results could be immediately generalized by considering sophisticated EoSs for the ultra-dense sector and constrain possible scenarios which might rule out or not twin stars. In a future study, we will consider the relation of the observables obtained in this work with the ones given by gravitational-wave data~\cite{Passamonti:2005cz,Passamonti:2007tm}.
 
\begin{acknowledgments}
This work was partially supported by INCT-FNA (Process No. 464898/2014-5). V.P.G. and L.L. acknowledges support from CNPq, CAPES (Finance Code 001), and FAPERGS. V.P.G. was partially supported by the CAS President's International Fellowship Initiative (Grant No.  2021VMA0019). J.C.J. acknowledges support from FAPESP (Processes No. 2020/07791-7 and No. 2018/24720-6). 
\end{acknowledgments}


\end{document}